# Electromagnetic theory of turbulent acceleration of parallel flow and momentum conservation

Shuitao Peng and Lu Wang[*]

State Key Laboratory of Advanced Electromagnetic Engineering and Technology, School of Electrical and Electronic Engineering, Huazhong University of Science and Technology, Wuhan, Hubei 430074, China

*E-mail: luwang@hust.edu.cn

## Abstract

Intrinsic flow in plasma physics is a long-standing puzzle, since it is difficult to understand its origin without contradiction to momentum conservation in conventional wisdom. It is proved that the electromagnetic turbulent acceleration as a candidate for intrinsic parallel flow generation driven by pressure gradient along the total magnetic field line does not contradict momentum conservation. The conserved quantity corresponding to axial symmetry is the total gyrocenter parallel canonical momentum carried by both species or the total gyrocenter parallel momentum including the ion gyrocenter kinematic momentum and electromagnetic fields momentum, but not the ion kinematic momentum, or even the ion parallel flow. A conservation equation of total parallel momentum including the ion particles' kinematic momentum and electromagnetic fields momentum is also presented.

## I. INTRODUCTION

Plasma flow generation and momentum transport are of great interest in laboratory [1], space [2], and astrophysical plasmas [3, 4]. Many instances of intrinsic flow generation and faster momentum transport than classical dissipation are attributed to arise from wave turbulence [5]. This type of problem has considerable overlap with the physics of turbulent magnetic dynamo [6]. Intrinsic plasma parallel flow (spontaneous flow without external momentum input) and anomalous momentum transport have been observed in magnetic confinement device [7]. It is well known that plasma rotation is beneficial for tokamak operation and performance via stabilizing macroscopic MHD instabilities such as resistive wall modes and neoclassical tearing mode and suppressing micro-turbulence. Understanding the generation mechanism for intrinsic flow [8] is especially important to the next generation devices, which cannot be adequately penetrated by conventional neutral beam injection, and so cannot achieve sufficient drive torque.

There has been intensive research in intrinsic flow, turbulent momentum transport and momentum conservation based on gyrokinetic theory. The divergence of residual stress which is a non-diffusive and non-convective component of the turbulent Reynolds stress is thought to be the origin of the intrinsic flow. The intrinsic stress force driven by electromagnetic turbulence has also been done [9, 10]. The canonical momentum conservation is obtained by summing gyrocenter momentum density over species and using gyrokinetic quasi-neutrality [11-16]. Ref. [11] also discussed collisonless momentum exchange between waves and particles. In Ref. [12], the momentum source due to radial current of gyrocenter can be written as time evolution of the field toroidal momentum density. In Refs. [17, 18], the turbulent acceleration for parallel flow cannot be written as a divergence of stress, and so acts as a local source or sink of parallel flow. However, there is still a frequently asked question: whether the turbulent acceleration for parallel flow contradicts momentum conservation or not.

In this paper, we prove that the turbulent acceleration as a candidate for the origin of intrinsic parallel flow does not contradict momentum conservation. The turbulent acceleration of parallel flow driven by ion pressure gradient along the total magnetic field (including background magnetic field and fluctuating radial magnetic field) cannot be written as a divergence of stress, so it acts as a local source/sink of parallel flow. We find that the conserved quantity corresponding to axial symmetry is the total gyrocenter canonical momentum density (summing over both species) or the total momentum including ion gyrocenter kinematic momentum and electromagnetic fields momentum, but not the ion kinematic momentum, or even ion flow. We also present a conservation equation of the ion particles' kinematic momentum plus electromagnetic fields momentum. Therefore, the turbulent acceleration of parallel flow does not conflict with momentum conservation.

The remainder of this paper is organized as follows. In Sec. II, we present the derivation of the mean parallel flow velocity evolution equation in two separated approaches. In Sec. III, momentum conservation equations are given. Finally, we summarize our work and discuss its possible implications for experiments in Sec. IV.

## II. PARALLEL FLOW VELOCITY EVULUTION EQUATION AND TURBULENT ACCELERATION

We start from the conservative form of the nonlinear electromagnetic gyrokinetic equation [19],

$$\frac{\partial (FB_\parallel^*)}{\partial t} + \nabla \cdot \left(\frac{d\overline{R}}{dt} FB_\parallel^*\right) + \frac{\partial}{\partial \bar{v}_\parallel} \left(\frac{d\bar{v}_\parallel}{dt} FB_\parallel^*\right) = 0, \tag{1}$$

with gyrocenter equations of motion in the symplectic formulation, i.e., $\bar{v}_\parallel$ representation [20]

$$\frac{d\overline{R}}{dt} = \bar{v}_\parallel \widehat{\boldsymbol{b}}^* + \frac{c}{eB} \widehat{\boldsymbol{b}} \times e\nabla \langle\langle \delta\phi_{\text{gc}} \rangle\rangle, \tag{2}$$

and

$$\frac{d\bar{v}_\parallel}{dt} = -\frac{e}{m_i}\left(\widehat{\boldsymbol{b}}^* \cdot \nabla \langle\langle \delta\phi_{\text{gc}} \rangle\rangle + \frac{1}{c}\frac{\partial \langle\langle \delta A_{\parallel\text{gc}} \rangle\rangle}{\partial t}\right). \tag{3}$$

Here, uniform equilibrium magnetic field is assumed. Only the shear component of magnetic perturbation, i.e., $\delta A_{\parallel\text{gc}}$, is considered. $m_i$ is the ion mass, $\delta\phi_{\text{gc}}$ is the gyrocenter electric potential fluctuation, and $\langle\langle \cdots \rangle\rangle$ denotes gyroaveraging. We denote $\langle\langle \delta A_{\parallel\text{gc}} \rangle\rangle = \delta A_\parallel$ and $\langle\langle \delta\phi_{\text{gc}} \rangle\rangle = \delta\phi$ for simplicity. In this paper, the index $\parallel$ refers to components parallel to the equilibrium magnetic field, and the index $\perp$ refers to components perpendicular to the equilibrium magnetic field. $F = F(\overline{R}, \bar{\mu}, \bar{v}_\parallel, t)$ is the ion gyrocenter distribution function with $\bar{\mu}$ being the gyrocenter magnetic moment, $B_\parallel^* = \widehat{\boldsymbol{b}} \cdot \boldsymbol{B}^* = B$ is the Jacobian of the transformation from the particle phase space to the gyrocenter phase space with $\boldsymbol{B}^* = \boldsymbol{B} + \delta\boldsymbol{B}_\perp$, $\delta\boldsymbol{B}_\perp = -\widehat{\boldsymbol{b}} \times \nabla \delta A_\parallel$. $\widehat{\boldsymbol{b}}^* = \widehat{\boldsymbol{b}} + \delta\widehat{\boldsymbol{B}}_\perp$ with $\delta\widehat{\boldsymbol{B}}_\perp = \frac{\delta \boldsymbol{B}_\perp}{B}$. Then, we can obtain the evolution equation of the ion gyrocenter density $\bar{n}_i \equiv (2\pi/m_i) \int d\bar{\mu} d\bar{v}_\parallel FB_\parallel^*$ by taking the zeroth order moment of the nonlinear electromagnetic gyrokinetic equation,

$$\frac{\partial \bar{n}_i}{\partial t} + \nabla \cdot \left[\left(\overline{U}_\parallel \widehat{\boldsymbol{b}}^* + \delta \mathbf{v}_{E\times B}\right)\bar{n}_i\right] = 0, \tag{4}$$

where $\bar{n}_i \overline{U}_\parallel \equiv (2\pi/m_i) \int d\bar{\mu} d\bar{v}_\parallel FB_\parallel^* \bar{v}_\parallel$ is the ion gyrocenter parallel kinematic momentum density per ion mass. By taking the first order moment, the ion gyrocenter parallel momentum density equation is written as

$$\frac{\partial(m_i\bar{n}_i\bar{U}_\parallel)}{\partial t} + \nabla \cdot \left(\widehat{\boldsymbol{b}}^* \bar{P}_{\parallel i} + \delta\mathbf{v}_{E\times B} m_i \bar{n}_i \bar{U}_\parallel\right) = -e\left(\widehat{\boldsymbol{b}}^* \cdot \nabla\delta\phi + \frac{1}{c}\frac{\partial \delta A_\parallel}{\partial t}\right)\bar{n}_i. \tag{5}$$

Here, $\bar{P}_{\parallel i} = 2\pi \int d\bar{\mu} d\bar{v}_\parallel F B_\parallel^* m_i \bar{v}_\parallel^2 = \bar{n}_i \bar{T}_i$ is the parallel ion gyrocenter pressure and $\frac{\bar{P}_{\parallel i}}{m_i} \gg \bar{n}_i \bar{U}_\parallel^2$ is assumed, $\delta\mathbf{v}_{E\times B} = \frac{c\widehat{\boldsymbol{b}}\times\nabla\delta\phi}{B}$ is the fluctuating $\boldsymbol{E}\times\boldsymbol{B}$ drift velocity. The first term in divergence includes an off-diagonal pressure tensor term, $\delta\widehat{\boldsymbol{B}}_\perp \bar{P}_{\parallel i}$. On the right hand side (RHS) of Eq. (5), it is the electromagnetic version of turbulent parallel momentum source due to electric field (including both electrostatic and inductive components) along the total magnetic field.

The ion parallel flow velocity is more experimentally measured quantity than the ion parallel momentum density. Firstly, multiplying Eq. (4) by $\bar{U}_\parallel$ and subtracting it from Eq. (5), we can obtain the ion gyrocenter parallel flow velocity equation

$$\frac{\partial \bar{U}_\parallel}{\partial t} + \nabla \bullet (\delta\mathbf{v}_{E\times B}\bar{U}_\parallel) = -\frac{e}{m_i}\left(\widehat{\boldsymbol{b}}^*\cdot\nabla\delta\phi + \frac{1}{c}\frac{\partial\delta A_\parallel}{\partial t}\right) - \frac{1}{m_i n_0}\widehat{\boldsymbol{b}}^*\cdot\nabla\bar{P}_{\parallel i}. \tag{6}$$

Note that in conventional tokamaks where toroidal magnetic field is much larger than poloidal magnetic field (i.e., $B_\varphi \gg B_\theta$), the ion toroidal flow velocity can be approximated by the ion gyrocenter parallel flow velocity [21] since $\bar{U}_\parallel = \bar{\boldsymbol{U}}\cdot\widehat{\boldsymbol{b}} = (\bar{U}_\varphi\widehat{\boldsymbol{e}}_\varphi + \bar{U}_\theta\widehat{\boldsymbol{e}}_\theta)\cdot\frac{(B_\varphi\widehat{\boldsymbol{e}}_\varphi + B_\theta\widehat{\boldsymbol{e}}_\theta)}{B} = \bar{U}_\varphi\frac{B_\varphi}{B} + \bar{U}_\theta\frac{B_\theta}{B} \cong \bar{U}_\varphi$. It is also noted that the ion gyrocenter parallel flow velocity can be approximated by the ion parallel flow velocity in particle space for a long wave length limit. This is the reason why we calculate the gyrocenter ion parallel flow velocity equation. And the mean field theory of the ion gyrocenter parallel flow velocity can be obtained by taking flux average of Eq. (6),

$$\frac{\partial\langle\bar{U}_\parallel\rangle}{\partial t} + \nabla\cdot\langle\delta\mathbf{v}_{E\times B,r}\delta\bar{U}_\parallel\rangle + \frac{1}{m_i n_0}\nabla\cdot\langle\delta\widehat{\boldsymbol{B}}_r\delta\bar{P}_{\parallel i}\rangle + \frac{e}{m_i}\nabla\cdot\langle\delta\widehat{\boldsymbol{B}}_r\delta\phi\rangle$$

$$= \frac{1}{m_i n_0}\left[\langle\delta\widehat{n}\widehat{\boldsymbol{b}}\cdot\nabla\delta\bar{P}_{\parallel i}\rangle + \langle\delta\widehat{n}\delta\widehat{\boldsymbol{B}}_r\rangle\cdot\nabla\bar{P}_{\parallel i}\right]. \tag{7}$$

This is the main result of this paper. Here, $\bar{U}_\parallel = \langle\bar{U}_\parallel\rangle + \delta\bar{U}_\parallel$ and $\delta\widehat{n} = \frac{\delta\bar{n}_i}{n_0}$ with $\bar{n}_i = \langle \bar{n}_i \rangle + \delta \bar{n}_i = n_0 + \delta \bar{n}_i$, $\delta \widehat{\boldsymbol{B}}_r = \frac{\delta B_r}{B}$. The electric field along the background field vanishes by taking a flux average, while the component along the fluctuating magnetic field can be written as a divergence of cross Maxwell stress, i.e., $\frac{e}{m_i} \nabla \cdot \langle \delta \widehat{\boldsymbol{B}}_r \delta \phi \rangle$ with $\nabla \cdot \delta \widehat{\boldsymbol{B}}_\perp = 0$ being used. The kinetic stress, $\langle \delta \widehat{\boldsymbol{B}}_r \delta \bar{P}_{\parallel i} \rangle$ is somewhat analogous to the kinetic dynamo physics where it is a transport process of electron parallel momentum (current) [22]. In MST, kinetic stress is experimentally identified to drive the intrinsic parallel plasma flow [23]. In DIII-D, the fluid Reynolds stress driven by electrostatic turbulence cannot explain the observed rotation profile, and Maxwell stress and kinetic stress are speculated as the candidates for the disagreement [24]. Both cross Maxwell stress and kinetic stress enter the parallel flow equation via their divergence. In this sense, they are surface force, which are similar to the usual Reynolds stress. In contrast, the turbulent acceleration on the RHS of Eq. (7) cannot be recast into a divergence, and is an effective volume-force. The turbulent acceleration driven by the parallel gradient of ion pressure fluctuation, $\frac{1}{m_i n_0} \langle \delta \hat{n} \widehat{\boldsymbol{b}} \cdot \nabla \delta \bar{P}_{\parallel i} \rangle$ was investigated for electrostatic turbulence in Refs. [17, 18]. The other one is directly related to magnetic fluctuation, $\frac{1}{m_i n_0} \langle \delta \hat{n} \delta \widehat{\boldsymbol{B}}_r \rangle \cdot \nabla \bar{P}_{\parallel i}$, which is driven by the equilibrium ion pressure gradient along $\delta \widehat{\boldsymbol{B}}_r$. Combining these two terms, the total electromagnetic turbulent acceleration can be written as the correlation between density fluctuation and perturbed pressure gradient along the total magnetic field, $\frac{1}{m_i n_0} \langle \delta \hat{n} \, \delta (\widehat{\boldsymbol{b}}^* \cdot \nabla \bar{P}_{\parallel i}) \rangle$. Electromagnetic effects may be important in H-mode pedestal [25].

An alternative method to derive the mean parallel flow is taking flux average of the gyrocenter parallel momentum density first, and then decoupling flow velocity and density as follows

$$\frac{\partial \langle \bar{U}_\parallel \rangle}{\partial t} = \frac{1}{n_0} \left[ \frac{\partial \langle \bar{n}_i \bar{U}_\parallel \rangle}{\partial t} - \langle \bar{U}_\parallel \rangle \frac{\partial n_0}{\partial t} - \langle \delta \bar{U}_\parallel \frac{\partial}{\partial t} \delta \bar{n}_i \rangle - \langle \delta \bar{n}_i \frac{\partial}{\partial t} \delta \bar{U}_\parallel \rangle \right], \qquad (8)$$

Taking flux average of Eqs. (5) and (4), we can obtain the mean gyroceneter parallel momentum density equation

$$\frac{\partial \langle \bar{n}_i \bar{U}_\parallel \rangle}{\partial t} = -\nabla \cdot \langle \delta v_{E \times B, r} \delta (\bar{n}_i \bar{U}_\parallel) \rangle - \frac{1}{m_i} \nabla \cdot \langle \delta \widehat{\boldsymbol{B}}_r \delta \bar{P}_{\parallel i} \rangle - \frac{e}{m_i} n_0 \nabla \cdot \langle \delta \widehat{\boldsymbol{B}}_r \delta \phi \rangle$$

$$-\frac{e}{m_i}\langle\delta\bar{n}_i\left(\widehat{\boldsymbol{b}}\cdot\nabla\delta\phi+\frac{1}{c}\frac{\partial\delta A_\parallel}{\partial t}\right)\rangle, \qquad (9)$$

and mean density equation

$$\frac{\partial n_0}{\partial t}=-\nabla\cdot\langle\delta v_{E\times B,r}\delta\bar{n}_i\rangle-\nabla\cdot\langle\delta\widehat{\boldsymbol{B}}_r\delta(\bar{n}_i\bar{U}_\parallel)\rangle. \qquad (10)$$

Linearizing density equation, Eq. (4), and parallel flow velocity equation, Eq. (6), and taking cross correlation, we can obtain

$$\langle\delta\bar{U}_\parallel\frac{\partial}{\partial t}\delta\bar{n}_i\rangle=-\langle\delta\bar{U}_\parallel\nabla\cdot(\delta\mathbf{v}_{E\times B}n_0)\rangle-\langle\delta\bar{U}_\parallel\nabla\cdot(n_0\langle\bar{U}_\parallel\rangle\delta\widehat{\boldsymbol{B}}_\perp)\rangle$$

$$-\langle\delta\bar{U}_\parallel\nabla\cdot\left(\delta(\bar{n}_i\bar{U}_\parallel)\widehat{\boldsymbol{b}}\right)\rangle, \qquad (11)$$

$$\langle\delta\bar{n}_i\frac{\partial}{\partial t}\delta\bar{U}_\parallel\rangle=-\langle\delta\bar{n}_i\nabla\cdot(\delta\mathbf{v}_{E\times B}\langle\bar{U}_\parallel\rangle)\rangle-\frac{e}{m_i}\langle\delta\bar{n}_i\left(\widehat{\boldsymbol{b}}\cdot\nabla\delta\phi+\frac{1}{c}\frac{\partial\delta A_\parallel}{\partial t}\right)\rangle$$

$$-\frac{1}{m_in_0}\left[\langle\delta\hat{n}\widehat{\boldsymbol{b}}\cdot\nabla\delta\bar{P}_{\parallel i}\rangle+\langle\delta\hat{n}\delta\widehat{\boldsymbol{B}}_r\rangle\cdot\nabla\bar{P}_{\parallel i}\right]. \qquad (12)$$

Putting Eqs. (9)-(12) into Eq. (8) yields

$$\frac{\partial\langle\bar{U}_\parallel\rangle}{\partial t}+\nabla\cdot\langle\delta v_{E\times B,r}\delta\bar{U}_\parallel\rangle+\frac{1}{m_in_0}\nabla\cdot\langle\delta\widehat{\boldsymbol{B}}_r\delta P_{\parallel i}\rangle+\frac{e}{m_i}\nabla\cdot\langle\delta\widehat{\boldsymbol{B}}_r\delta\phi\rangle$$

$$=\frac{1}{m_in_0}\left[\langle\delta\hat{n}\widehat{\boldsymbol{b}}\cdot\nabla\delta\bar{P}_{\parallel i}\rangle+\langle\delta\hat{n}\delta\widehat{\boldsymbol{B}}_r\rangle\cdot\nabla\bar{P}_{\parallel i}\right]. \qquad (13)$$

Here, some terms have been neglected with assumptions $\frac{\bar{P}_{\parallel i}}{m_i}\gg\bar{n}_i\bar{U}_\parallel^2$ and $\frac{\delta\bar{P}_{\parallel i}}{m_i}\gg\delta\bar{n}_i\bar{U}_\parallel^2,\bar{n}_i\bar{U}_\parallel\delta\bar{U}_\parallel$. We note that the mean parallel flow equations (7) and (13) are the same regardless of the sequence of taking flux average and decoupling the parallel flow from parallel momentum density. The presence of turbulent acceleration does not result from regrouping terms. From the later method, we can see the turbulent acceleration due to parallel fluctuating electric field is cancelled, but perturbed pressure gradient along the total magnetic field driven acceleration is kept.

### III. MOMENTUM CONSERVATION EQUATIONS

Taking summation of Eq. (5) over both species and using quasi-neutrality condition, we obtain

$$\frac{\partial}{\partial t}\sum_s(m_s\bar{n}_s\bar{U}_{\|s}) + \nabla \cdot \sum_s(\widehat{\boldsymbol{b}}^*\bar{P}_{\|s} + \delta\mathbf{v}_{E\times B}m_s\bar{n}_s\bar{U}_{\|s}) = en_{\text{pol}}\left(\widehat{\boldsymbol{b}}^* \cdot \nabla\delta\phi + \frac{1}{c}\frac{\partial \delta A_\|}{\partial t}\right). \quad (14)$$

Here, $s$ includes ions and electrons, $-\sum_s e_s \bar{n}_s = en_{\text{pol}} = \nabla_\perp \cdot \left(\frac{c^2 m_i n_0}{B^2}\nabla_\perp \delta\phi\right)$ is the ion polarization charge density with finite Larmor radius effects for electrons being ignored. $n_{\text{pol}} = -\nabla \cdot \boldsymbol{P}_{\text{gy}}$ is the ion polarization density, where $\boldsymbol{P}_{\text{gy}} = -\frac{c^2 m_i n_0}{B^2}\nabla_\perp \delta\phi = \int d^3\mathbf{v}\, F\boldsymbol{\pi}_{\text{gy}}$ is the polarization vector with $\boldsymbol{\pi}_{\text{gy}} = -\frac{c^2 m_i}{B^2}\nabla_\perp \delta\phi$ being the gyrocenter electric-dipole moment [26]. We also ignores the contribution from the magnetic flutter term $-\nabla_\perp \cdot \left(\frac{c^2 m_i n_0}{B^2}\frac{\bar{U}_{\|i}}{c}\nabla_\perp \delta A_\|\right)$ to the ion polarization density because of the condition $\frac{\bar{U}_{\|i}}{c}\nabla_\perp \delta A_\| \ll \nabla_\perp \delta\phi$. This is consistent with the gyrokinetic ordering $\frac{v_\|}{c}\frac{e\delta A_\|}{T_i} \sim \frac{e\delta\phi}{T_i}$ [26] and the ordering $\bar{U}_{\|i} \ll v_\| \sim v_{thi} = \sqrt{\frac{T_i}{m_i}}$ we take in this work. Moreover, the ordinary kinematic momentum density carried by electrons can be neglected due to $m_e \ll m_i$. Then, taking flux average of Eq. (14) yields

$$\frac{\partial \langle m_i \bar{n}_i \bar{U}_{\|i}\rangle}{\partial t} + \nabla_r \cdot \langle \delta\mathbf{v}_{E\times B,r}\delta(m_i\bar{n}_i\bar{U}_{\|i}) + \delta\widehat{\boldsymbol{B}}_r \delta\bar{P}\rangle$$

$$= e\left\langle n_{\text{pol}}\left(\widehat{\boldsymbol{b}} \cdot \nabla\delta\phi - \frac{\widehat{\boldsymbol{b}}\times\nabla\delta A_\|}{B} \cdot \nabla\delta\phi + \frac{1}{c}\frac{\partial \delta A_\|}{\partial t}\right)\right\rangle. \quad (15)$$

Here, $\delta\bar{P} = \delta\bar{P}_{\|i} + \delta\bar{P}_{\|e}$ is the total gyrocenter parallel pressure. Taking summation of the ion gyrocenter density continuity equation and the electron density continuity equation, we can obtain $\frac{\partial n_{pol}}{\partial t} + \nabla \cdot \langle \delta\mathbf{v}_{E\times B,r}n_{pol}\rangle = 0$ [15]. This is equivalent to the polarization charge conservation law $\frac{\partial \rho_{pol}}{\partial t} + \nabla \cdot \mathbf{J}_{\text{pol}} = 0$ with $\rho_{pol} = en_{pol} = \nabla_\perp \cdot \left(\frac{c^2 m_i n_0}{B^2}\nabla_\perp \delta\phi\right)$, and $\mathbf{J}_{\text{pol}} = \frac{\partial \boldsymbol{P}_{\text{gy}}}{\partial t} = \frac{\partial}{\partial t}\left(-\frac{c^2 m_i n_0}{B^2}\nabla_\perp \delta\phi\right)$ is the polarization current [26]. In other words, the ion polarization density flux induced by $\boldsymbol{E} \times \boldsymbol{B}$ drift velocity is equivalent to the ion density flux induced by polarization drift velocity. Then, after some algebra, we can obtain

$$\frac{\partial}{\partial t}\sum_s\langle\bar{p}_{\|s}\rangle + \nabla_r \cdot \langle\delta\mathbf{v}_{E\times B,r}\delta(m_i\bar{n}_i\bar{U}_{\|i}) + \delta\widehat{\boldsymbol{B}}_r\delta\bar{P}\rangle - \nabla_r \cdot \langle\frac{c^2 m_i n_0}{B^2}\nabla_r\delta\phi\widehat{\boldsymbol{b}}\cdot\nabla\delta\phi\rangle = 0,$$
$$(16)$$

where $\sum_s \langle \bar{p}_{\parallel s} \rangle = \langle m_i \bar{n}_i \bar{U}_{\parallel i} - n_{pol} \frac{e}{c} \delta A_\parallel \rangle$ with $\bar{p}_{\parallel s} = m_s \bar{n}_s \bar{U}_{\parallel s} + \bar{n}_s \frac{e_s}{c} \delta A_\parallel$ being the gyrocenter canonical parallel momentum of one species including the ordinary gyrocenter kinematic momentum and the gyrocenter magnetic vector momentum. The total gyrocenter magnetic vector momentum $-n_{pol} \frac{e}{c} \delta A_\parallel$ results from the ion polarization density. We have conserved the total gyrocenter canonical parallel momentum carried by both ions and electrons. The pieces under the divergence are radial flux of the ion gyrocenter parallel kinematic momentum, total kinetic stress and electric Maxwell stress, respectively.

Using gyrocenter density continuity equation, we can obtain the relationship between total gyrocenter magnetic vector momentum and electromagnetic fields momentum

$$\frac{\partial \langle -n_{pol} \frac{e}{c} \delta A_\parallel \rangle}{\partial t} = \frac{\partial \langle m_i n_0 \frac{c^2}{B^2} (\delta E \times \delta B)_\parallel \rangle}{\partial t} - \nabla \cdot \langle \frac{c^2 m_i n_0}{B^2} \nabla_r \delta \phi \frac{1}{c} \frac{\partial \delta A_\parallel}{\partial t} \rangle. \qquad (17)$$

On the RHS, the piece under the time derivative is the parallel projection of fluctuating electromagnetic fields momentum density [27], $\boldsymbol{g}_F = \delta \boldsymbol{D} \times \delta \boldsymbol{B}$, where $\delta \boldsymbol{D} = \epsilon \delta \boldsymbol{E}$ is the electric displacement field fluctuation, with $\epsilon = m_i n_0 \frac{c^2}{B^2}$ being the permittivity. This means $-n_{pol} \frac{e}{c} \delta A_\parallel$ is essentially related to the electromagnetic fields momentum. The other one under the divergence is Maxwell stress corresponding to radial electrostatic and parallel inductive electric fields. Thus, Eq. (16) can be rewritten as

$$\frac{\partial \langle m_i \bar{n}_i \bar{U}_{\parallel i} + m_i n_0 \frac{c^2}{B^2} (\delta E \times \delta B)_\parallel \rangle}{\partial t} + \nabla_r \cdot \langle \delta v_{E \times B, r} \delta(m_i \bar{n}_i \bar{U}_{\parallel i}) + \delta \widehat{\boldsymbol{B}}_r \delta \bar{P} \rangle$$

$$- \nabla_r \cdot \langle \frac{c^2 m_i n_0}{B^2} \nabla_r \delta \phi \left( \widehat{\boldsymbol{b}} \cdot \nabla \delta \phi + \frac{1}{c} \frac{\partial \delta A_\parallel}{\partial t} \right) \rangle = 0. \qquad (18)$$

This is another format of momentum conservation. The quantity under the time derivative is the total gyrocenter parallel momentum density which includes the ion gyrocenter kinematic momentum and electromagnetic fields momentum. The pieces under the divergence are radial flux of the ion gyrocenter kinematic momentum, total kinetic stress, and Maxwell stress including parallel inductive electric field. Similar gyrocenter toroidal momentum conservation equation was derived for electrostatic turbulence in Ref. [12].

Now, we discuss the momentum conservation in particle space. By pullback transformation, the ion particles' kinematic momentum can be written as

$$p_{\parallel i} = m_i \bar{n}_i \bar{U}_{\parallel i} + \frac{m_i}{e} \nabla \times \mathbf{M}_{\text{gy}}$$

$$\approx m_i \bar{n}_i \bar{U}_{\parallel i} + \nabla_\perp \cdot \left( p_{\parallel i} \frac{m_i c^2}{eB^2} \nabla_\perp \delta\phi \right), \quad (19)$$

where $\mathbf{M}_{\text{gy}} = \int d^3\mathbf{v}\, F \left( \boldsymbol{\mu}_{\text{gy}} + \boldsymbol{\pi}_{\text{gy}} \times \frac{\bar{v}_\parallel}{c}\hat{\boldsymbol{b}} \right)$ with $\boldsymbol{\mu}_{\text{gy}}$ being the gyrocenter magnetic-dipole moment. The second term on the RHS of Eq. (19) is the ion magnetization momentum induced by the moving electric-dipole moment contribution $\boldsymbol{\pi}_{\text{gy}} \times \frac{\bar{v}_\parallel}{c}\hat{\boldsymbol{b}}$ which corresponds to the ion magnetization current related to $\boldsymbol{\pi}_{\text{gy}}$ in Refs. [20, 26, 27]. The ion magnetization momentum induced by $\boldsymbol{\mu}_{\text{gy}}$ corresponds to the ion magnetization current related to $\boldsymbol{\mu}_{\text{gy}}$ [20, 26], and has been neglected for the isotropic ion pressure condition. Substituting Eq. (19) into Eq. (18), we can obtain

$$\frac{\partial \langle p_{\parallel i} + m_i n_0 \frac{c^2}{B^2}(\delta\mathbf{E}\times\delta\mathbf{B})_\parallel \rangle}{\partial t} - \nabla_r \cdot \langle \frac{m_i c^2}{eB^2} \frac{\partial}{\partial t}(\nabla_r \delta\phi \delta p_{\parallel i}) \rangle + \nabla_r \cdot \langle \delta v_{E\times B, r} \delta p_{\parallel i} + \delta\hat{\boldsymbol{B}}_r \delta P \rangle - \nabla_r \cdot \langle \frac{c^2 m_i n_0}{B^2} \nabla_r \delta\phi \left( \hat{\boldsymbol{b}} \cdot \nabla \delta\phi + \frac{1}{c}\frac{\partial \delta A_\parallel}{\partial t} \right) \rangle = 0. \quad (20)$$

Here, $\delta P = \delta P_{\parallel i} + \delta P_{\parallel e}$ is the total parallel pressure in particle space. The difference between $\delta P$ and $\delta\bar{P}$ is a higher order term, and thus can be neglected for the quasi-linear theory. Eq. (20) is the conservation equation of total parallel momentum in particle space, which includes the ion particles' kinematic momentum and electromagnetic fields momentum. The second term is related to momentum flux induced by the polarization drift. Different from the total gyrocenter parallel momentum, there is only one type of momentum conservation in particle space. This is because there is no net particles' magnetic vector momentum after cancellation by summing over both species.

We have shown that the above two gyrocenter momentum conservation equations are equivalent. The conserved quantity can be either the total gyrocenter parallel canonical momentum carried by both species or the total gyrocenter parallel momentum including the ion gyrocenter kinematic momentum and electromagnetic fields momentum but not the ion parallel flow velocity. What's more, we have demonstrated that the total parallel momentum in particle space including the ion particles' kinematic momentum and electromagnetic fields momentum is also conserved. Therefore, turbulent source or sink of the mean ion parallel flow *does not* imply that the parallel momentum is not conserved.

## IV. SUMMARY AND DISCUSSIONS

In this paper, the mean parallel flow velocity equation for electromagnetic turbulence is derived. We prove that the turbulent acceleration of parallel flow does not contradict momentum conservation. We point out that the conserved quantity corresponding to axial symmetry is the total gyrocenter parallel canonical momentum carried by both species or the total gyrocenter parallel momentum including the ion gyrocenter kinematic momentum and electromagnetic fields momentum, and a conservation equation of the total parallel momentum in particle space is also derived. There have been no simulation works or experimental measurements on turbulent acceleration yet. The methods for testing the theory of turbulent acceleration by gyrokinetic simulation proposed in Ref. [17] may be also applicable to electromagnetic turbulence. The observed parallel flow in MST [23] might agree better with the total intrinsic flow drive including both the kinetic stress force and the corresponding turbulent acceleration than with the kinetic stress force only. Therefore, we suggest experimentalists to explore the evidence for turbulent acceleration.

## ACKNOWLEDEMENTS

We are thankful to P. H. Diamond, H. Jhang and W. X. Ding for useful discussions. This work was supported by the NSFC Grant Nos. 11305071 and 11675059, and the Ministry of Science and technology of China under Contract No. 2013GB112002.